# Trustworthy Artificial Intelligence for Cyber Threat Analysis

Shuangbao Paul Wang[1(B)] and Paul A. Mullin[2]

[1] Department of Computer Science, Morgan State University, Baltimore, MD, USA
paul@nisti.org
[2] Softrams, LLC, Leesburg, VA, USA

**Abstract.** Artificial Intelligence brings innovations into the society. However, bias and unethical exist in many algorithms that make the applications less trustworthy. Threats hunting algorithms based on machine learning have shown great advantage over classical methods. Reinforcement learning models are getting more accurate for identifying not only signature-based but also behavior-based threats. Quantum mechanics brings a new dimension in improving classification speed with exponential advantage. In this research, we developed a machine learning-based cyber threat detection and assessment tool. It uses two-stage (unsupervised and supervised learning) analyzing method on 822,226 log data recorded from a web server on AWS cloud. The results show the algorithm has the ability to identify the threats with high confidence.

## 1 Overview

The advancement of Artificial Intelligence (AI) has accelerated the adoption and integration of innovations into many frontiers including automobiles, banking, insurances, etc. Machine learning can reveal a lot of things that human beings can hardly find out. On the theoretical side, Machine Learning (ML) and deep learning algorithms are able to not only analyze data efficiently but also accumulate the knowledge gained from previous learning. The ML models are getting improved from time to time with new feed-in data to the neural networks under reinforcement learning. On the other hand, the accuracy or even the correctness of the AI/ML algorithms could be affected by many factors, from algorithm, data, to prejudicial, or even intentional. As a result, AI/ML applications need to be not only accurate, robust, reliable, transparent, non-biased, and accountable but also deemed trustworthy and are able to adapt, recover, reconfigure in response to challenges. To achieve this goal, foundational study about trust and trustworthiness is vital. Use-inspired research can bring new discoveries into commercialization to benefit the society.

Cyber threat analysis involves billions of data either live or recorded from servers, firewalls, intrusion detection and prevention systems, and network devices [25,27]. Signature-based analyzing method is effective only if the attack vectors are pre-defined and already stored in the knowledge database. Hence, if threat actors change their behavior, it would be hard to capture. To identify the abnormal behaviors, threat analytics systems need to not only examine the irregular patterns at certain time but also observe the behavior changes comparing with the normal states.

ML algorithms can be very helpful in identifying cyber threats with reinforcement learning models. They can be classified as:

– **Unsupervised Learning**. A clustering model attempts to find groups, similarities, and relationships within unlabelled data, Fig. 1 illustrates a unsupervised learning model.

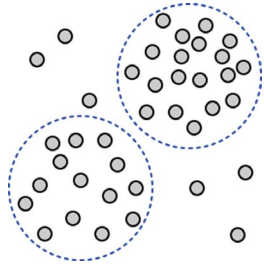

**Fig. 1.** Unsupervised Learning

– **Supervised Learning**. A classification model to identify how input variables contribute to the categorization of data points. Figure 2 depicts a supervised learning model.

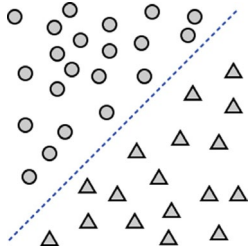

**Fig. 2.** Supervised Learning

– **Semi-Supervised Learning**. A classification model falls between supervised learning and unsupervised learning by combining a small amount of labeled data with a large amount of unlabeled data during training.

- **Reinforcement Learning**. Reinforcement learning is characterized by a continuous loop where an agent interacts with an environment and measures the consequences of its actions. Figure 3 shows a reinforcement learning model. [1]

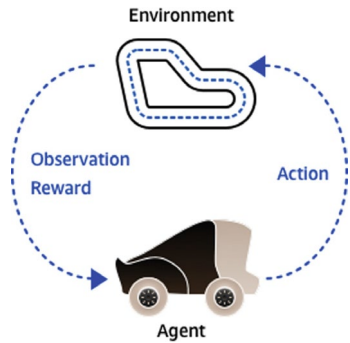

Fig. 3. Reinforcement Learning

– **Transfer Learning**. Transfer learning stores knowledge gained while solving one problem and applies it to a different but related problem.

Public sector organizations operate external-facing applications with broad multi-task user interfaces. There are significant investments made in human-centered design, but new opportunities exist to enhance personalization of the interfaces through AI/ML to reduce burden on the public users in navigation and task performance. The problem involves leveraging information from user roles, prior behavior, schedules of required activities, and other characteristics to predict the intended reasons why users are entering a system at any given time, and where they plan to navigate and work. This information would be used to facilitate the users in those navigations. Tracking and log data from servers on AWS or Google Analytics can be used for exploration in the development of a methodology. More Machine Learning complex navigation analytics can be explored using behavior analysis method and visualized through business intelligence dashboards.

## 1.1 Literature Review

Artificial intelligence revolutionizes industry and everyday life in many aspects. The Deep Blue computer can defeat the greatest human chess player in the world. The autonomous vehicles such as Tesla can drive on the road without human interactions. Machine learning can reveal a lot of things that human beings can hardly find out. By analyzing music using IBM Watson AI, people can learn the mode of songs and hence discovered that the majority of songs from 60s to now are in the mode of "sadness".

---

[1] Fig. 1, 2, 3 image source: [3].

In cybersecurity, AI/ML is used to deep inspect the packets, analyze the network activities, and discover abnormal behaviors.

Sagar et al. conducted a survey of cybersecurity using artificial intelligence [7]. It discusses the need for applying neural networks and machine learning algorithms in cybersecurity.

Mittu et al. proposed a way to use machine learning to detect advanced persistence threats (APT) [17]. The approach can address APT that can cause damages to information systems and cloud computing platforms.

Mohana et al. proposed a methodology to use genetic algorithms and neural networks to better safe guard data [4]. A key produced by a neural network is said to be stronger for encryption and decryption.

With a grant from the National Science Foundation (NSF), Wang and Kelly developed a video data analytics tool that can penetrate into videos to "understand" the context of the video and the language spoken [26].

Kumbar proposed a fuzzy system for pattern recognition and data mining [14]. It is effective in fighting phishing attacks by identifying malware.

Using Natural Language Processing (NLP), Wang developed an approach that can identify issues with cybersecurity policies in financial processing process so financial banking companies can comply with PCI/DSS industry standards.

Harini used intelligent agent to reduce or prevent distributed denial of service (DDoS) attacks [21]. An expert system is used to identify malicious codes to prevent being installed in the target systems.

With a grant from National Security Agency (NSA), Wang and his team developed an intelligent system for cybersecurity curriculum development. The system is able to develop training and curricula following the National Initiative of Cybersecurity Education (NICE) framework.

Dilrmaghani et al. provide an overview of the existing threats that violate security and privacy within AI/ML algorithms [8].

Gupta et al. studied quantum machine learning that uses quantum computation in artificial intelligence and deep neural networks. They proposed a quantum neuron layer aiming to speed up the classification process [13].

Mohanty et al. surveyed quantum machine learning algorithms and quantum AI applications [18].

Edwards and Rawat conducted a survey on quantum adversarial machine learning by adding a small noise that leads to classifier to predict to a different result [10]. By depolarization, noise reduction and adversarial training, the system can reduce the risk of adversarial attacks.

Ding at al. proposed a quantum support vector machine (SVM) algorithm that can achieve exponential speedup for least squares SVM (LS-SVM) [9]. The experiments show it has the advantages in solving problems such as face recognition and signal processing.

Ablayev et al. provided a review of quantum methods for machine learning problems [2]. In quantum tools, it lists the fundamentals of qubits, quantum registers, and quantum states, tools on quantum search algorithms. In quantum classification algorithms, it introduces several classification problems that can be accelerated by using quantum algorithms.

Wang & Sakk conducted a survey of overviews, foundations, and speedups in quantum algorithms [28]. It provides a detailed discussion on period finding, the key of quantum advantage to factor large numbers.

### 1.2 Bias in Existing AI/ML Algorithms

People hope the neutrality in AI/ML algorithms. Unfortunately, bias does exist. *Washington Post* published an article that "credit scores are supposed to be race-neural. That's impossible". *Forbes* published a report that "self-driving card are more likely to recognize white pedestrians than black pedestrians, resulting in decreased safety for darker-skinned individuals". A *Nature* paper reports that "a major healthcare company used an algorithm that deemed black patients less worthy of critical healthcare than others with similar medical conditions". *Associate Press* published an article that "financial technology companies have been shown to discriminate against black and latinx households via higher mortgage interest rates".

In general, bias can be classified into the following categories:

- **Algorithm Bias**. Bias as a result of inaccurate algorithms are used.
- **Data Bias**. Bias due to incorrectly sample the data for training that are not reflect the whole data set.
- **Prejudicial Bias**. Feeding model with prejudicial knowledge, for example, "nurses are female".
- **Measurement Bias**. Bias as a result of incorrect measurement.
- **Intentional Bias**. People embed unjust or discriminatory rules in the AI/ML models.

Trustworthy AI/ML is to discover those bias and build robust AI/ML algorithms that is trustworthy. For example, a Tesla with autopilot could crash onto a fire truck, which is hardly possible even for the worst human drivers. This shows that the nine "eyes" (radar system) under artificial intelligence are still inferior than two eyes of human intelligence. To be trustworthy, Tesla cars need to be trained with reliable data that can "filter" the "noise" caused by the emergency light flashing, which changes the images of a fire truck.

Effort has been put in countering the AI bias. Obaidat et al. uses random sampling on images with a convolutional neural network (CNN) [20]. They tested using the Fashion-MNIST dataset that consists of 70,000 images, 60,000 for training and 10,000 for testing with an accuracy of 87.8%.

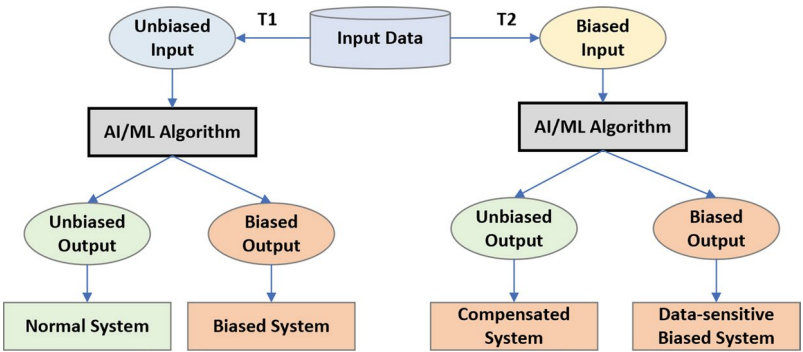

**Fig. 4.** Two-tier analyzing method for rating bias in AI/ML algorithms

Bernagozzi et al. at IBM conducted a survey that reveals the presence of bias in chatbots and online language translators using two-tier method to rate bias [5]. The two-tier method is illustrated in Fig. 4.

Lohia et al. at IBM proposed a framework that can detect bias to improve fairness [15]. The algorithms detects individual bias and then post-processing for bias mitigations.

## 2   Using Adversarial ML Model to Discover Bias

Adversarial ML model is commonly used to attack ML algorithms in a way that the models would function abnormally. McAfee once attacked Tesla's system by adding a strip to a speed limit sign that fooled the car to drive 50 mph over the speed limit. A "stealth" ware technology with adversarial pattern on glasses and clothes can fool facial recognition systems. Adversarial attacks can be classified into three categories:

- **Evasion**. An attack uses steganography and other technologies to obfuscate the textual content.
- **Poisoning**. An attack to contaminate the training data.
- **Model stealing**. An attack to consider the target as a black box and try to extract data from the model.

By introducing quantum neuron layers, there is a potential to speed up the classification process with an acceptable error rate.

### 2.1   Notation of Bias and Mitigation

Consider a supervised classification problem with features $X \in \mathsf{X}$, categorical protected attributes $D \in \mathsf{D}$, and categorical labels $Y \in \mathsf{Y}$. We are given a set of training samples $\{(x_1, d_1, y_1), ..., (x_n, d_n, y_n)\}$ and would like to learn a classifier $\hat{y}$ : X × D → Y. For ease of exposition, we will only consider a scalar binary protected attribute, i.e. D = {0, 1}. The value $d = 1$ is set to correspond to the *privileged* group (e.g. whites in the United States in criminal justice application) and $d = 0$ to *unprivileged* group (e.g. blacks). The value $y = 1$ is set to correspond to a *favorable* outcome. Based on the context, we may also deal with probabilistic binary classifiers with continuous output scores $\hat{y}_S \in [0, 1]$ that are thresholded to {0, 1}.

One definition of individual bias is as follows. Sample $i$ has individual bias if $\hat{y}(x_i, d = 0) = \hat{y}(x_i, d = 1)$. Let $b_i = I[\hat{y}(x_i, d = 0) = \hat{y}(x_i, d = 1)]$, where $I[\cdot]$ is an indicator function. The individual bias score, $b_{S,i} = \hat{y}_S(x + i, d = 1) - \hat{y}_S(x_i, d = 0)$, is a soft version of $b_i$. To compute an individual bias summary statistic, we take the average of $b_i$ across test samples.

One notion of group fairness known as *disparate impact* is defined as follows [15]. There is disparate impact if the division of the expected values

$$\frac{E[\hat{y}(X, D)|D = 0]}{E[\hat{y}(X, D)|D = 1]} \quad (1)$$

is less than $1 - E$ or greater than $(1 - E)^{-1}$, where a common value of $E$ is 0.2.

The test procedure usually divided into two steps: 1) determine whether there are any trials of individual bias, 2) discover the found bias against all samples.

Mitigation can be performed by changing the label outputs of the classifier $\hat{y}_i$ to other labels $\breve{y} \in Y$.

Zhang et al. use federated learning (FL) in privacy-aware distributed machine learning application [29]. Experiments show it can reduce the discrimination index of all demographic groups by 13.2% to 69.4% with the COMPAS dataset.

## 3 AI Bias and ML-Based Threat Analytics

Artificial intelligence uses data to train models and uses an inference engine to draw a conclusion or predict the outcome. The overall architecture can be shown as Fig. 5.

Su et al. utilize open source testing and analysis tools Hping3, Scapy to simulate DDoS flood and import the data collected from those simulated attacks into Splunk to identify possible attacks [22]. The Splunk machine learning toolkit extends the capabilities of Splunk.

Ngoc et al. proposed an early warning approach to counter APT. It analyzes the APT target using log analysis techniques [19].

*Counterfit* is an AI security risk assessment package developed at Microsoft. The open-source tool helps companies conduct AI risk assessment to ensure the AI/ML algorithms are non-bias, reliable, and trustworthy [16].

## 4 Programming and Experiments

Detecting web attacks using machine learning is an area that has drawn attention and requires continuous research and development. This project analyzes 822,226 log records from a healthcare IT company's web login page in a 5 h time span. After cleaning and pre-processing the data, the algorithm detected records that could potentially be attacks. It then calculated the likelihood (of attacks) based on the abnormal behaviors.

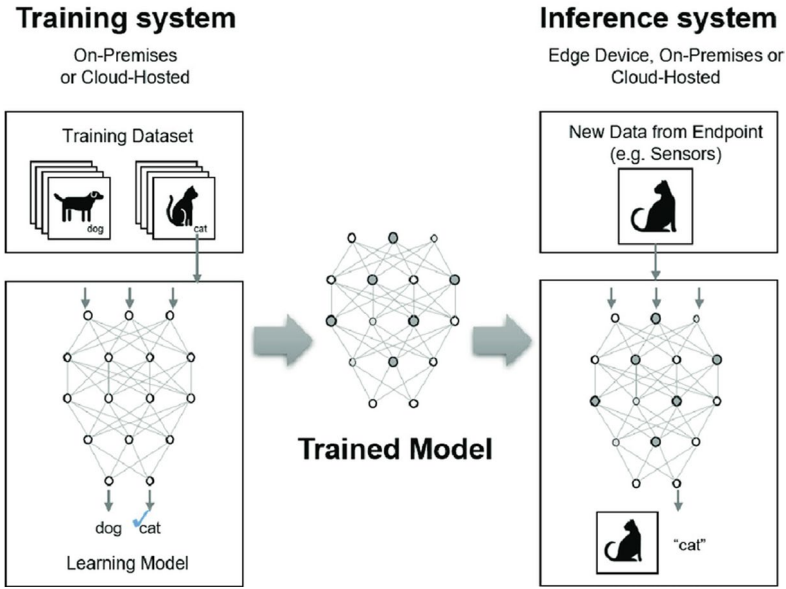

**Fig. 5.** Machine learning architecture.

The main strategy is to use unsupervised learning for better understanding the distribution of the input data. Supervised learning is then applied for further classification and generating predictions. As a result, the model learns how to predict/classify on output from new inputs. Reinforcement learning learns from experiences over time. The algorithm can be improved with more data feed into the system.

The application first loads the input data into a Pandas dataframe, then removes features that are not of interests in detecting attacks. Next, data are "compressed" from 800,000+ to around 40,000 by combining the records that have the same source and destination ip addresses in the same unit time period.

The higher the compression rate, the more the duplications in the dataset. This improved the efficiency of machine learning process. Unsupervised machine learning is applied to the dataset using K-means clustering. The output three clusters are labeled as not-suspicious, suspicious, and transitional area, as shown in Fig. 6.

The pre-processed data are then splitted into 0.66/0.33 for training/testing and further analyzing the likelihood of each response's abnormal behaviors. Using results (three clusters) from the unsupervised learning as a supervisor, the algorithm continues apply supervised machine learning to discover the threats.

In addition to areas that are considered "confident" or "no confident", the transition (in yellow) area is further analyzed using k-mean clustering to separate into 2 clusters, labeled as "more suspicious" and "less suspicious", as shown in Fig. 7. The "more suspicious" tags are then added into the suspicious activity dataset. By doing so it ensures the machine does not miss any responses that get filtered out during analyzing process but is still suspected having abnormal behaviors. The likelihood of the suspicion is calculated based on the percentage of "attacks" over the maximum responses per second.

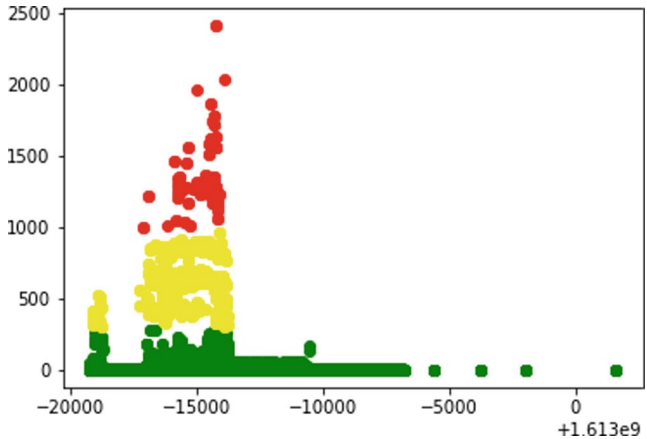

**Fig. 6.** Classification using unsupervised learning

An attack for a general log-in page is defined as considerable number of visits, responses, callbacks in a short period of time. Thus, pre-processing the data by combining each duplicating responses per second helps determine the number of responses or visits that stands out.

The application can be improved by feeding into more and rich data so risks associated with human behaviors can be identified.

The 3–2 two-tier classification technique helps narrowing down the suspicious activities. If the k-means clustering is applied only once with 2 clusters, the uncertain groups of dataset would possibly be wrong. Therefore, creating a transition area in the middle of two certainties helps detecting the potential attacks that could be missed.

The result is saved into result.csv and all detected attacks are saved in the suspicious_activity.csv. A list of references for this Python application can be found in [1,6,11,12,23,24].

The team is conducting quantum-inspired machine learning research to identify cyber threats in data and in AI/ML tools. It is expected that the new findings will further improve the speed and accuracy of identifying cyber threats.

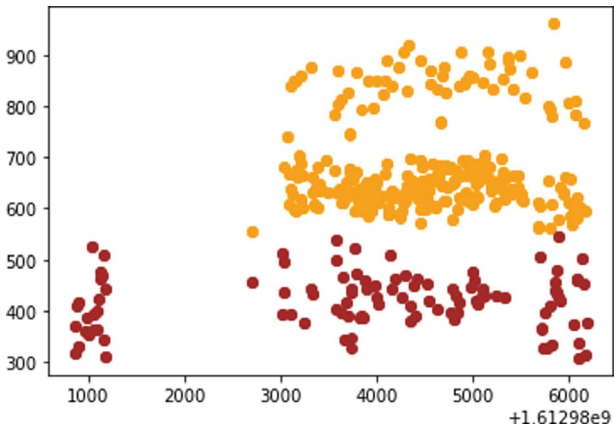

**Fig. 7.** Classification using supervised learning